\documentclass[11pt]{article}

\usepackage[round]{natbib} 
\usepackage{graphics, graphicx, hyperref, url, amsmath, amsthm, amsfonts, enumerate, epstopdf, subfigure, booktabs, float}
\hypersetup{pdfstartview={FitH}}

     \def\etal{{\itshape et al.}}
\def\beqn{\begin{eqnarray}} \def\eeqn{\end{eqnarray}}

\newcommand{\R}{\mathbb{R}}

\newcommand{\PD}{{\mbox{\it PD}}}

\newcommand{\bec}{\begin{center}}
\newcommand{\enc}{\end{center}}
\newcommand{\bee}{\begin{eqnarray*}}
\newcommand{\ene}{\end{eqnarray*}}
\newcommand{\beq}{\begin{equation}}
\newcommand{\eeq}{\end{equation}}

\begin{document}

\title{\bf Exactly computing bivariate projection depth contours and median
\thanks{Corresponding author's email: zuo@msu.edu, tel: 001-517-432-5413.}}

\author {{Xiaohui Liu$^{a,b}$,  Yijun Zuo$^{*b}$,  Zhizhong Wang$^a$}\\
         {\em\footnotesize $^a$ School of Mathematics Science and Computing Technology, Central South University, Hunan 410083, China}\\
         {\em\footnotesize $^b$ Department of Statistics and Probability, Michigan State University, East Lansing, MI 48823, USA}\\
         }
\date{}
\maketitle

\begin{center}
\begin{minipage} {12cm}{ %\bec {\sc Abstract  } \enc
\vspace*{-7mm}
\small \textbf{Abstract}.
\small Among their competitors, projection depth and its induced estimators are very favorable because they can enjoy very high breakdown point robustness without having to pay the price of low efficiency, meanwhile providing a promising center-outward ordering of multi-dimensional data. However, their further applications have been severely hindered due to their computational challenge in practice. In this paper, we derive a simple form of the projection depth function, when $(\mu, \sigma)$ = (Med, MAD). This simple form enables us to extend the existing result of point-wise exact computation of projection depth (PD) of \cite{ZuoLai2011} to depth contours and median for bivariate data.
%and provide a rigorous proof that the computation involved for PD in higher dimensions just needs to consider
%finitely many projections when outlyingness function involves (Med, MAD).
%there is no need to calculate the supremum over an infinite
%number of directions for the case where outlyingness function involves (Med, MAD) in higher dimensions. Furthermore, We found
%that these finite directions will not change in the whole process of depth computation as long as the data %cloud is fixed.
%Furthermore, we are also able to construct an exact algorithm for computing bivariate projection depth median.
%}
\vspace{1mm}

\vspace{1mm}

}
\end{minipage}
\end{center}

\vskip 0.1 in
\section{Introduction}
\paragraph{}
\vskip 0.1 in
\label{Introduction}

To generalize order-related univariate statistical methods, depth functions have emerged as powerful tools for nonparametric multivariate analysis with the ability to provide a center-outward ordering of the multivariate observations. Points deep inside a data cloud get higher depth and those on the outskirts receive lower depth. Such depth induced ordering enables one to develop favorable new robust estimators of multivariate location and scatter matrix. Since Tukey's introduction \citep{Tuk1975}, depth functions have gained much attention in the last two decades. Numerous depth notations have been introduced. To name a few, halfspace depth \citep{Tuk1975}, simplicial depth \citep{Liu1990}, regression depth \citep{RouHub1999}, projection depth \citep{Liu1992, ZuoSer2000, Zuo2003}.
\medskip

\cite{ZuoSer2000} and \cite{Zuo2003} found that
%Zuo and Serfling (2000) and Zuo (2003) found that
among all the examined depth notions, projection depth
%using formula $PD(x, F) = 1 / (1 + O(x, F))$ with $O(x, F)$
%being the outlyingness function \citep{ZuoSer2000, Zuo2003},
is one of the favorite, enjoying very desirable properties.
%Fox example, (1) it satisfies all the four desirable properties of depth function mentioned in \cite{ZuoSer2000},
%i.e., Affine Invariance, Maximality at Center, Monotonicity relative the Deepest point, Vanishing at infinity, (2)
%both of its empirical and population version are continuous in $x$ and they are always positive, (3), what's more %important,
Furthermore, projection depth induced robust estimators, such as projection depth weighted means and median, can
possess a very high breakdown point as well as high relative efficiency with appropriate choices of univariate location
and scale estimators, serving as very favorable alternatives to the regular mean \citep{Zuo2003, Zuoetal2004}. In fact,
projection depth weighted means include as a special case the famous Donoho-Stahel estimator \citep{Sta1981,
Don1982, Tyl1994, Maretal1995, Zuoetal2004}, the latter is the first constructed location estimator  in high dimensions
enjoying high breakdown point robustness and affine equivariance, while the projection depth median has the highest
breakdown point among all the existing affine equivariant multivariate location estimators \citep{Zuo2003}.
\medskip

However, the further prevalence of projection depth and its induced estimators is severely hindered by their computational intensity and intimidation. The computation of projection depth seems intractable since it involves
%The key point lies in the fact that the computation of projection depth involves the
supremum over infinitely many direction vectors. There were only approximating algorithms in the last three decades until \cite{ZuoLai2011}, in which they proved that there is no need to calculate the supremum over infinitely many direction vectors in the bivariate data when the outlyingness function uses the very popular choice (Med, MAD) as the univariate location and scale pair. An exact algorithm for projection depth and its weighted mean, i.e. Donoho-Stahel estimator, was also constructed in that paper. 
\medskip

In the current paper, we further generalize their idea to the higher dimensional cases by utilizing linear fractional functionals programming \citep{Swa1962}. That is, we find that, with the choice of (Med, MAD), we only need to calculate the supremum over a finite number of direction vectors for $p \ge 2$. Furthermore, these direction vectors are $x$-free and depend only on the data cloud. Therefore, we derive a simple form of the projection depth function, and are able to compute the bivariate projection depth contours and median very conveniently through linear programming based on the procedure of \cite{ZuoLai2011}. It is found that sample projection depth contours are polyhedral under some mild conditions. Furthermore, it is noteworthy that the computational methods discussed in this paper have no limitation on the dimension $p$, and therefore could possible be implemented to spaces with $p > 2$, as well as to the modified projection depth \citep{Sim2011} in a more general multidimensional regression context. 
%Furthermore, we found that, once the data cloud is fixed, these finite directions remain the same in the process of
%depth calculations over different point. Based on these observations, 
The corresponding programs are available from the authors (zuo@msu.edu or csuliuxh912@gmail.com).
\medskip

The rest of the paper is organized as follows. Section \ref{Definition} provides the definitions of the projection depth contour and projection median. Section \ref{TheMainIdea} presents the main idea of how to get a simple form of the projection depth function. Section \ref{EABD} discusses the exact computational issue of the projection depth contour and projection median by linear programming. While some numerical examples are given in Section \ref{numAna}.

\vskip 0.1 in
\section{Definition}
\paragraph{}
\vskip 0.1 in \label{Definition}

%For a given distribution $F$ on $\R^1$, let $\mu(F)$ be some translation equivariant and scale invariant robust univariate location measurements and  $\sigma(F)$ be some translation invariant and scale equivariant robust univariate scale measurements. 
For a given distribution $F$ on $\R^1$, let $\mu(F)$ be translation equivariant and scale invariant, and $\sigma(F)$ be translation invariant and scale equivariant. Define the outlyingness of a point $x\in R^{p}\ (p \ge 1)$ with respect to the distribution $F$ of the random variable $X\in R^{p}$ as (see \citep{Zuo2003} and references therein)
\begin{equation}
    O(x, F) = \sup_{\|u\|=1} |Q(u, x, F)|
    \label{Eqn001}
\end{equation}
where $Q(u, x, F) = (u^{\tau}x - \mu(F_{u})) / \sigma(F_{u})$, if $u^{\tau}x-\mu(F_u)=\sigma(F_u)=0$, then define $Q(u,
x, F)=0$. $F_{u}$ is the distribution of $u^{\tau}X$, which is the projection of $X$ onto the unit vector $u$.
\medskip

Throughout this paper, we select the very popular robust choice of $\mu$ and $\sigma$: the median (Med) and the median absolute deviation (MAD). Based on definition \eqref{Eqn001}, the projection depth of any given point $x$ with respective to $F$, $\PD(x, F)$, can then be defined as \citep{Liu1992, ZuoSer2000, Zuo2003}
$$
    \PD(x, F) = 1 / (1 + O(x, F))
$$

With the outlyingness function and projection depth function defined above, we then define the projection depth median
(PM) and contours (PC) as follows \citep{Zuo2003}
$$
    PM(F) = \mbox{arg} \max\limits_{x\in R^{p}} PD(x, F),
$$
$$
    PC(\alpha, F) = \left\{x \in R^{p} : PD(x, F) = \alpha\right\},
$$
where $0 < \alpha \leq \alpha^{*} = \sup_{x\in R^p}PD(x,F)$.
\medskip

For a given sample $\mathcal{X}^{n} = \{X_{1}, X_{2}, \cdots, X_{n}\}$ from $X$, let $F_{n}$ be the corresponding
empirical distribution. By simply replacing $F$ by $F_{n}$ in $PM(F)$ and $PC(\alpha, F)$, we can obtain their sample
version: $PM(F_{n})$ and $PC(\alpha, F_{n})$. Without confusion, we use $\mathcal{X}^{n}$ and $F_n$ interchangeably in what follows. Furthermore, by noting the fact that for the choice of (Med, MAD), $Q(u, x, \mathcal{X}^{n})$ in \eqref{Eqn001} is odd with respect to $u$, we drop the absolute value sign existing in definition \eqref{Eqn001}, and consider 
$$
    O(x, \mathcal{X}^{n}) = \sup\limits_{\|u\| = 1} Q(u, x, \mathcal{X}^{n})
$$
instead in what follows, where 
$$
Q(u, x, \mathcal{X}^{n}) = \frac{u^{\tau}x - \mbox{Med}
    (u^{\tau}\mathcal{X}^{n})}{\mbox{MAD}(u^{\tau}\mathcal{X}^{n})},
$$
where $u^{\tau}x$ denotes the projection of $x$ onto the unit vector $u$, and $u^{\tau}\mathcal{X}^{n} = \left\{u^{\tau}X_{1}, u^{\tau}X_{2}, \cdots, u^{\tau}X_{n}\right\}$. Let $Z_{(1)} \leq Z_{(2)} \leq, \cdots, \leq Z_{(n)}$ be the order statistics based on the univariate random variables $\mathcal{Z}^{n} = \{Z_{1}, Z_{2}, \cdots, Z_{n}\}$, then 
$$
    \mbox{Med}(\mathcal{Z}^{n}) = \frac{Z_{(\lfloor (n + 1) / 2\rfloor)} + Z_{(\lfloor (n + 2) / 2\rfloor)}}{2},
$$
$$
    \mbox{MAD}(\mathcal{Z}^{n}) = \mbox{Med}\{|Z_{i} - \mbox{Med}(Z^{n})|, i = 1, 2, \cdots, n\},
$$
where $\lfloor \cdot \rfloor$ is the floor function.
\medskip

%also write $PM(\mathcal{X}^{n})$ and $PC(\alpha, \mathcal{X}^{n})$ for $PM(F_{n})$ and
%$PC(\alpha, F_{n})$, respectively.

\vskip 0.1 in
\section{The main idea}
\paragraph{}
%\vskip -0.25 in \label{TheMainIdea}

%\subsection{Simple form of $O(x, \mathcal{X}^{n})$}
%\paragraph{}
\vskip 0.1 in \label{TheMainIdea}

Note that, for any given sample $\mathcal{X}^{n}$, the tasks of computing both $PM(\mathcal{X}^{n})$ and $PC(\alpha,
\mathcal{X}^{n})$ mainly involve $O(x, \mathcal{X}^{n})$, i.e.
$$
    PM(\mathcal{X}^{n}) = \mbox{arg} \min\limits_{x\in R^{p}} O(x, \mathcal{X}^{n}),
$$
$$
    PC(\alpha, \mathcal{X}^{n}) = \left\{x \in R^{p} : O(x, \mathcal{X}^{n}) = \beta\right\},
$$
where $\beta = 1 / \alpha - 1$. Thus, let's  first focus on the computation of $O(x, \mathcal{X}^{n})$. Without loss of generality, in what follows, we assume $\mathcal{X}^{n}$ to be in general position, which is commonly used in most existing literature; see for example \cite{DonGas1992}.
\medskip

By the idea of a circular sequence \citep{Ede1987} (see also \cite{Dyc2000, Cas2007}), for any given unit vector $v \in \mathcal{S} = \{u \in R^{p} : \|u\| = 1\}$, there must exist two permutations, say $(i_{1}, i_{2}, \cdots, i_{n})$ and $(j_{1}, j_{2}, \cdots, j_{n})$, of $(1, 2, \cdots, n)$ such that
$$
    v^{\tau}X_{i_{1}} \leq v^{\tau}X_{i_{2}} \leq \cdots \leq v^{\tau}X_{i_{n}},
$$
$$
    Y_{ j_{1}} \leq Y_{ j_{2}} \leq \cdots \leq Y_{ j_{n}},
$$
where $Y_{ j_{l}} = |v^{\tau}X_{j_{l}} - \mbox{Med}(v^{\tau}\mathcal{X}^{n})|,\ (1 \leq l \leq n)$. There is a small non-empty set $\mathcal{N}(v)\subset \mathcal{S}$ of $v$ such that these hold true for any $v \in
\mathcal{N}(v)$. This implies that the whole unit sphere $\mathcal{S}$ can be covered completely by at most $N$ (with order $O\big({n \choose p}^2\big)$) non-empty fragments $S_{k} = \{ v\in \mathcal{S}$: $v$ satisfy constraint conditions $\mathcal{Q}_{k}\}$ with
$\mathcal{Q}_{k}$ being
$$
\left\{
    \begin{array}{c}
        v^{\tau}  (X_{i_{k,2}} - X_{i_{k,1}}) \ge 0,\\
        v^{\tau}  (X_{i_{k,3}} - X_{i_{k,2}}) \ge 0,\\
        \vdots\\
        v^{\tau}  (X_{i_{k,n}} - X_{i_{k,n-1}}) \ge 0,\\
        Y_{j_{k,2}} - Y_{j_{k,1}} \ge 0,\\
        Y_{j_{k,3}} - Y_{j_{k,2}} \ge 0,\\
        \vdots \\
        Y_{j_{k,n}} - Y_{j_{k,n-1}} \ge 0,\\
        \|v\| = 1,
    \end{array}\right.
$$
for some fixed permutations $(i_{k, 1}, i_{k, 2}, \cdots, i_{k, n})$ and $(j_{k,1}, j_{k,2}, \cdots, j_{k,n})$ of $(1,\cdots, n)$,  where $1 \leq k \leq N$. %(When p=2, $\kappa(n)=O(n^2)$. See \cite{ZuoLai2011})
%with $s(n) = n(n - 1) / 2$.
%\emph{{ This should be $O\big(\big(n{n-1 \choose p-1}\big)^2\big)$ if you want to claim to extend the previous
%results to higher dimensions. Otherwise this $\kappa(n)$ holds just for $p=2$}}
\medskip

Note that different fragments $S_{k}\ (k = 1, 2, \cdots, N)$ are connected and overlapped with each other only on the
boundaries. Thus, to calculate $O(x, \mathcal{X}^{n})$, it is sufficient to calculate
$$
    O(x, \mathcal{X}^{n}) = \max_{1 \leq k \leq N} O_{k}(x, \mathcal{X}^{n})
$$
with
\begin{equation}
    O_{k}(x, \mathcal{X}^{n}) = \sup\limits_{u \in S_{k}} Q(u, x,
    \mathcal{X}^{n}).
    \label{TmpEq003}
\end{equation}
\medskip

Furthermore, from the definition and property of $S_{k}$, it is easy to see that, for any $u \in S_{k}$, the  outlyingness function $Q(u, x, \mathcal{X}^{n})$ can be simplified to
\begin{equation}
    Q(u, x, \mathcal{X}^{n}) = \frac{u^{\tau}(x - X_{i_{k, m}})}{|u^{\tau}X_{j_{k, m}} - u^{\tau}X_{i_{k, m}}|},
    \label{TmpEq}
\end{equation}
if $n$ is odd with $m = (n + 1) / 2$, otherwise
\begin{equation}
    Q(u, x, \mathcal{X}^{n}) = \frac{2u^{\tau}\left(x - X_{k, a}\right)}
    {\left|u^{\tau}\left(X_{j_{k, m}} - X_{k, a}\right)\right| +
    \left|u^{\tau}\left(X_{j_{k, m^{*}}} - X_{k, a}\right)\right|},
    \label{TemEqn002}
\end{equation}
with $m = n / 2$, $m^{*} = m + 1$ and $X_{k, a} = (X_{i_{k, m}} + X_{i_{k, m^{*}}}) / 2$.
\bigskip

\textbf{Remark 1.} Based on the assumption that $\mathcal{X}^{n}$ are in general position, the denominators in the above two formulas will not be 0 for all $u \in S_{k}$, since they are actually equal to MAD($u^{\tau}\mathcal{X}^{n}$), and greater than 0 under such an assumption when $n \ge 2p$; see the proof of Theorem 3.4 in \cite{Zuo2003}.
\bigskip

By \eqref{TmpEq} and \eqref{TemEqn002}, we obtain the following proposition.
\bigskip

\textbf{Proposition 1}. Assume $\mathcal{X}^{n}$ is in general position. Then for any given $k$ $(1\leq k\leq N)$, the
optimization problem \eqref{TmpEq003} is equivalent to
\begin{equation}
    O_{k}(x, \mathcal{X}^{n}) = \sup_{z} \frac{c_{k}^{\tau} z}{d_{k}^{\tau} z},
    \label{LinearPro}
\end{equation}
subject to
\begin{equation}
    A_{k} z \ge 0
    \label{ConstConditions}
\end{equation}
where $c_{k}, d_{k}$ and $A_{k}$ will be specified in the Appendix. Here $\textbf{b} \ge 0$ means that $\textbf{b}$ is component-wise non-negative if $\textbf{b}$ is a vector, i.e. for any component $\textbf{b}_{i}$, we have $\textbf{b}_{i} \ge 0$.
\medskip

\eqref{LinearPro} with constraint conditions \eqref{ConstConditions} is typically a linear fractional functionals
programming problem. By theorem 1 of \cite{Swa1962} (see also \cite{Sim2011} (p. 950) for more general discussion), it is easy to show that the maximum of $c_{k}^{\tau} z / d_{k}^{\tau} z$ will only occur at the basic feasible solution of \eqref{ConstConditions}. Note that the number of fragments $S_{k}$ is limited (at most $N$). Thus, we have
\bigskip

\textbf{Theorem 1.} Suppose that the choice of location and scale measures of projection depth function is the pair (Med, MAD). Then the number of direction vectors needed to compute the projection depth exactly is finite. Furthermore, these direction vectors only depend on the data cloud $\mathcal{X}^{n}$.
\bigskip

\textbf{Remark 2.} The idea of dividing the unit sphere $\mathcal{S}$ into fragments $S_{k}$ by applying Med and MAD sequences was first used in \cite{ZuoLai2011} for computing the bivariate projection depth; see also \cite{PS2011b, PS2011c} for other similar applications. Here we extend the result of \cite{ZuoLai2011} to $R^p$ ($p\geq2$). That is, one could compute PD in $R^p$ exactly by only considering a finite number of direction vectors. Furthermore, the $x$-free property of these direction vectors can bring convenience to the computation of PD$(x,\mathcal{X}^{n})$ for any $x$, since we only need to search the direction vectors once.
%However, it is still not clear in their paper that
%there is whether or not a need to calculate the supremum over infinite directions to obtain the exact depth
%value for higher dimensional data. The answer is negative, i.e., the number is finite, according to theorem
%1.\vskip 0.1 in

\vskip 0.1 in
\section{Exact computation of $PM(\mathcal{X}^{n})$ and $PC(\alpha, \mathcal{X}^{n})$}
\paragraph{}
\vskip 0.1 in \label{EABD}

From the discussion above, we can obtain the two following observations, namely, for any given $x$,
\begin{itemize}

\item{} the way to divide sphere $\mathcal{S}$ into fragments $S_{k}\ (k=1, 2, \cdots, N)$ is fixed, i.e.
$x$-free, as long as the data cloud $\mathcal{X}^{n}$ is fixed.

\item{} there is no need to calculate $Q(u, x, \mathcal{X}^{n})$ over an infinite number of direction vectors. It is enough to calculate it for, say, $u_{1}, u_{2}, \cdots, u_{M}$.
\end{itemize}

Based on the discussion and two observations above, we therefore can re-express the outlyingness function $O(x, \mathcal{X}^{n})$ as follows
$$
    O(x, \mathcal{X}^{n}) = \max \left\{\frac{u_{1}^{\tau}x - \text{Med}(u_{1}^{\tau}\mathcal{X}^{n})} {\text{MAD}(u_{1}^{\tau}\mathcal{X}^{n})}, \frac{u_{2}^{\tau}x - \text{Med}(u_{2}^{\tau}\mathcal{X}^{n})} {\text{MAD}(u_{2}^{\tau}\mathcal{X}^{n})}, \cdots, \frac{u_{M}^{\tau}x - \text{Med}(u_{M}^{\tau}\mathcal{X}^{n})} {\text{MAD}(u_{M}^{\tau}\mathcal{X}^{n})} \right\},
$$
where $\{ u_{i} \}_{i=1}^{M}$ are some $p$-dimensional vectors depending only on the data cloud $\mathcal{X}^{n}$. For the sake of convenience, hereafter we write $g_{i}(x) = \textbf{a}_{i}^{\tau}x - b_{i}$ $(i = 1, 2, \cdots, M)$, where $\textbf{a}_{i} = \frac{1}{\text{MAD}(u_{i}^{\tau} \mathcal{X}^{n})} u_{i}$ and $b_{i} = \frac{\text{Med}(u_{i}^{\tau} \mathcal{X}^{n})}{\text{MAD}(u_{i}^{\tau} \mathcal{X}^{n})}$. 
\medskip

Obviously, $O(x, \mathcal{X}^{n}) = \max_{1\leq i\leq M}\{g_i(x)\}$ is in fact a piece-wise linear convex function with respect to $x$ for the given data cloud $\mathcal{X}^{n}$. Therefore, its minimizers can be found by using common linear programming methods by solving the problem 
\begin{eqnarray*}
\label{PM001}
	s = \min_{z}\ t
\end{eqnarray*}
subject to 
$$
	t \ge g_{i}(x),\ i = 1, 2, \cdots, M,
$$
where $z = (t, x^{\tau})^{\tau}$. This kind of problem can be solved by some common solver such as \textit{linprog.m} in Matlab. Let $z_{0} = (t_{0}, x_{0}^{\tau})^{\tau}$ be a final solution of this problem. Then, it is easy to show that $x_{0}$ is one of the deepest points with depth value $PD(x_{0}, \mathcal{X}^{n}) = 1 / (1 + t_{0}) = \alpha^{*}$.
\medskip

Given the nature of the maximum piece-wise linear convex function $\max_{1\leq i\leq M} g_{i}(x)$, there is either a single minimizer or a convex polyhedral set of minimizers. %Then there naturally comes a question: is $z_{0}$ the unique solution?
%\medskip
%The following theorem provides an answer to this question.
%{\bf I stop here and will work on the
%proof of the theorem, but you need to work from the very beginning to 3.2 and afterwards, make sure there are no typos
%and state clearly what you want say}
%\bigskip
%\textbf{Theorem 2.} Suppose all the data points $X_{i} \in R^{p}\ (p\ge 1)$ are in general position. Then the set of the maximizers of the projection depth $PD(x, \mathcal{X}^{n})$, i.e. the minimizers of $O(x, \mathcal{X}^{n})$, is convex and has no inner points.
%\bigskip
%This theorem coincides with the result of Theorem 2.3 of \cite{Zuo2003}, which considered the population version of the projection median. The fact that the set of the maximizers of $PD(x, \mathcal{X}^{n})$ has no inner point implies that this set must be either a single point or a linear segment when $p = 2$, and either a single point or a linear or planar segment when $p = 3$, etc. For convenience, one can choose the average of all the vertexes of $\{x: PD(x, \mathcal{X}^{n}) = \alpha^{*} = 1 / (1 + t_{0})\}$ as the computed projection median in practice.
%\medskip
Then there naturally comes a question, namely, after obtaining the value $\alpha^{*}$, how to get all of these vertexes? Note that the projection median is a specific case of the projection depth contour. Therefore, let's focus now on the computation of projection depth contours. For any given $0 < \alpha \leq \alpha^{*}$, the projection depth contour is the boundary of the projection depth region \citep{Zuo2003}, 
\begin{eqnarray*}
    PR(\alpha, \mathcal{X}^{n}) &=& \left\{x \in R^{p} : PD(x, \mathcal{X}^{n}) \ge \alpha \right\}\\
    & = & \left\{x \in R^{p} : O(x, \mathcal{X}^{n}) \leq \beta \right\}\\
    & = & \left\{x \in R^{p} : g_{i}(x) \leq \beta, i = 1, 2, \cdots, M\right\}.
\end{eqnarray*}

Typically, the regions constrained by linear inequalities such as 
\begin{eqnarray}
\label{LPPR001}
	g_{i}(x) \leq \beta, i = 1, 2, \cdots, M
\end{eqnarray}
are polytopes. Therefore, the boundary of $PR(\alpha, \mathcal{X}^{n})$, i.e. $PC(\alpha, \mathcal{X}^{n})$, could be easily found by employing procedures such as \textit{qhull} \citep{Baretal1996} based on the dual relationship between vertex and facet enumeration \citep{Breetal1998}; see also \cite{PS2011a}. In Matlab, these kinds of tasks can be fulfilled by the function \textit{con2vert.m}, which was developed by Michael Kleder, and can be downloaded from Matlab Central File Exchange.
\medskip

However, in many practical applications, the number $M$ (with order $O\big({n \choose p}^2\big)$) of the direction vectors may be very large. When $M$ is too large, it is difficult to obtain the boundary of the region formed by \eqref{LPPR001} by using some of the aforementioned procedures such as \textit{con2vert.m}, since they involve solving some large generalized inverse matrices. Therefore, it is important to eliminate some redundant constraints before computing $PC(\alpha, \mathcal{X}^{n})$ for too large $M$. 
\medskip

Note that, for any given $\alpha$ $(0 < \alpha \leq \alpha ^{*})$, the number of the non-redundant constraints in \eqref{LPPR001} is much small compared to $M$, which implies that numerous inequalities in \eqref{LPPR001} could be eliminated during the computation of the $\alpha$-contour. In fact, it is not difficult to show that 
$$
	PR(\alpha, \mathcal{X}^{n}) = \mathcal{C}_{1} \cap \mathcal{C}_{2} \cap \cdots \cap \mathcal{C}_{s},
$$
where $\mathcal{C}_{k} = \{x : g_{j}(x) \leq \beta, j = i_{k}, i_{k} + 1, \cdots, i_{k + 1}\}$ with $1 \leq k \leq s - 1$ and $1 = i_{1} < i_{2} < \cdots < i_{s} = M$, and that $\mathcal{C}_{k}$ is non-empty if $\min_{1 \leq l \leq s - 1} \{i_{l + 1} - i_{l}\} \gg p$. Then, when $M$ is large, a procedure for exactly computing $PC(\alpha, \mathcal{X}^{n})$ is: (1), to find the non-redundant constraints in each $\mathcal{C}_{i}$ at first, then (2), to use all of these non-redundant constrains together to compute $PC(\alpha, \mathcal{X}^{n})$.
\medskip

With the vertexes in hand, some common graphical packages could be employed to visualize these contours very easily in spaces of $p = 2, 3$. It is noteworthy that, although all the methods discussed above could possible be implemented to spaces with $p \ge 2$ theoretically, feasible exact algorithm for computing the projection depth exists now only in the bivariate data \citep{ZuoLai2011}. Therefore, we can only provide some exactly results about the bivariate projection depth contours and projection median in the current paper. All the direction vectors $\{u_{1}, \cdots, u_{M}\}$ are found by using the procedure of \cite{ZuoLai2011}. The $x$-free property of these vectors can be proved similarly by using the linear fractional functionals programming as Proposition 1. %Since the main goal of \cite{ZuoLai2011} is to find the angular regions $\mathcal{C}$ such that, within each sub-angular region, it holds that
%\begin{eqnarray*}
%	\text{Med}(u^{T}\mathcal{X}^{n}) = u^{T}X_{j} \quad \text{and}\quad \text{MAD}(u^{T}\mathcal{X}^{n}) = (|u^{T}X_{i} - u^{T}X_{j}| + |u^{T}X_{k} - u^{T}X_{j}|) / 2,
%\end{eqnarray*}
%for some $i, j\in \{1, \cdots, n\}$ and for any $u \in \mathcal{C}$, where $j$ is determined by the Med sequences and $i$ and $k$ by the MAD sequence. $i$ and $k$ are the same in the odd $n$ case. 

\vskip 0.1 in
\section{Numerical analysis}
\paragraph{}
\vskip 0.1 in \label{numAna}

In order to gain more insight into the sample version of projection depth contours and projection median, we construct some numerical examples in this section.

\vskip 0.1 in
\subsection{Simulation results}
\paragraph{}
\vskip 0.1 in \label{SimuResu}

To illustrate the robustness and the shape of the bivariate projection depth contours and median, we present two examples as follows. The data are mainly generated from the normal distribution, but contain a few outliers. 
\medskip

{\bf Example 1.} We first generate 60 samples $X = (X_{1}, X_{2})^{\tau}$ from the normal distribution $N(0, I_{2})$, and then disturb these samples randomly by replacing their first components by 6 with probability 0.05.
\medskip

%the model: $(1 - \varepsilon) X + \varepsilon (\delta_{(0, 6)}, X_{2})^{\tau}$ with $\varepsilon = 0.05$, where $X = (X_{1}, X_{2})^{\tau} \sim N(0, I_{2})$, $\delta_{(0, 6)}$ is a discrete random variable with probability 0.95 taking value $6$ and 0.05 taking value $0$.
%(\textbf{\textsl{what does your $\delta_{(0, 6)}$ here and below mean?}})
%\medskip

{\bf Example 2.} We first generate 400 samples $X = (X_{1}, X_{2})^{\tau}$ from the normal distribution $N(0, \Sigma_{0})$, and then disturb these samples randomly by replacing their first components by 6 with probability 0.10, where
%We general 400 samples from model: $(1 - \varepsilon) X + \varepsilon (\delta_{(0, 6)}, X_{2})^{\tau}$
%with $\varepsilon = 0.1$, where $X = (X_{1}, X_{2})^{\tau}$ $\sim N(0, \Sigma_{0})$ with
$$
    \Sigma_{0} = \left(
    \begin{array}{lr}
        1 & 0.5\\
        0.5 & 1
    \end{array}
    \right).
$$

\begin{figure}[H] %\vspace*{-7mm}
    \begin{center}
    \includegraphics[angle=0,width=3.5in]{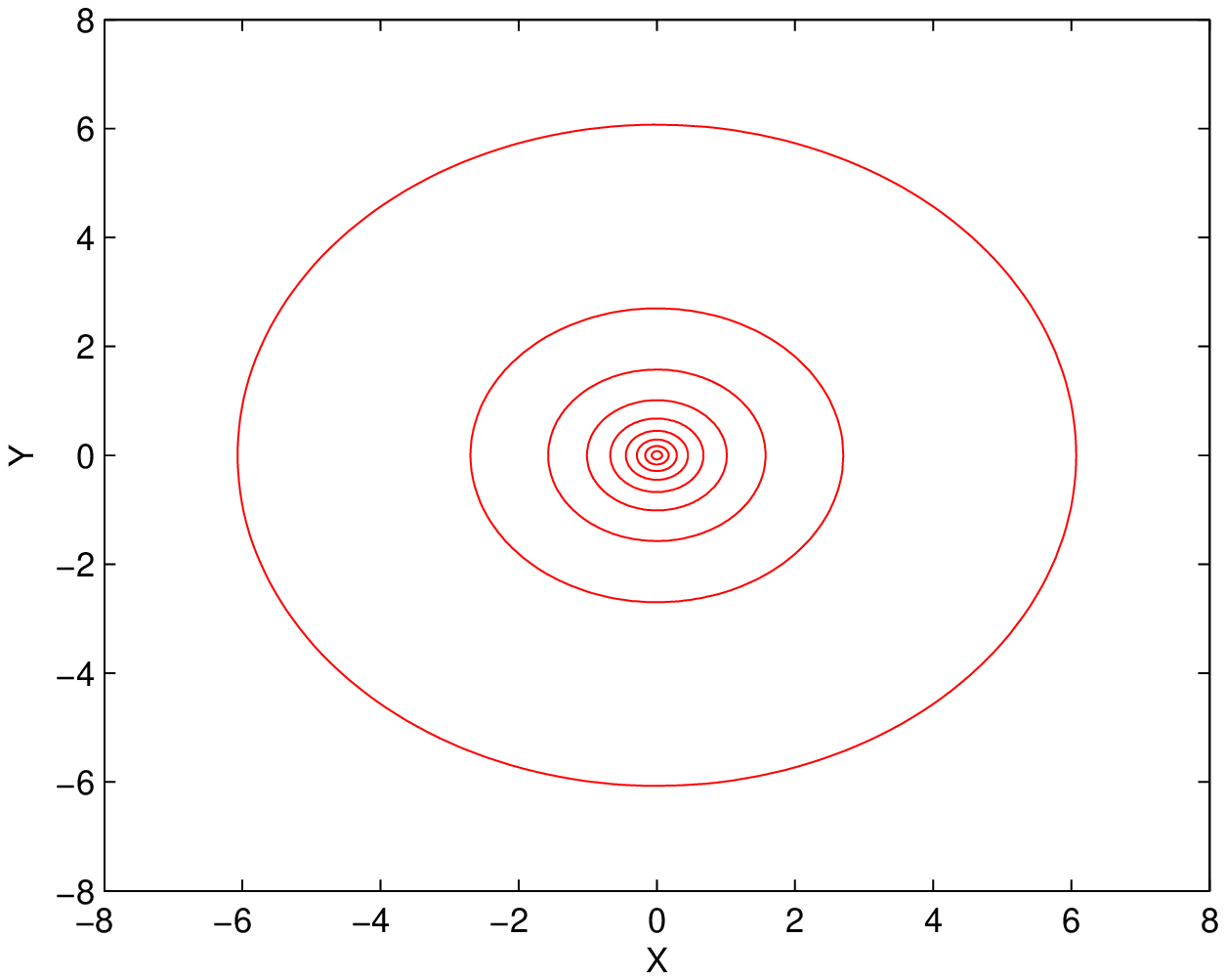}%
    \caption{Population version $PC(\alpha, X)$ for Example 1 with $\alpha = 0.1, 0.2, \cdots, 0.9$ (from out to center).}
    \label{Norm1PV60}
    \end{center}
    \begin{center}
    \includegraphics[angle=0,width=3.5in]{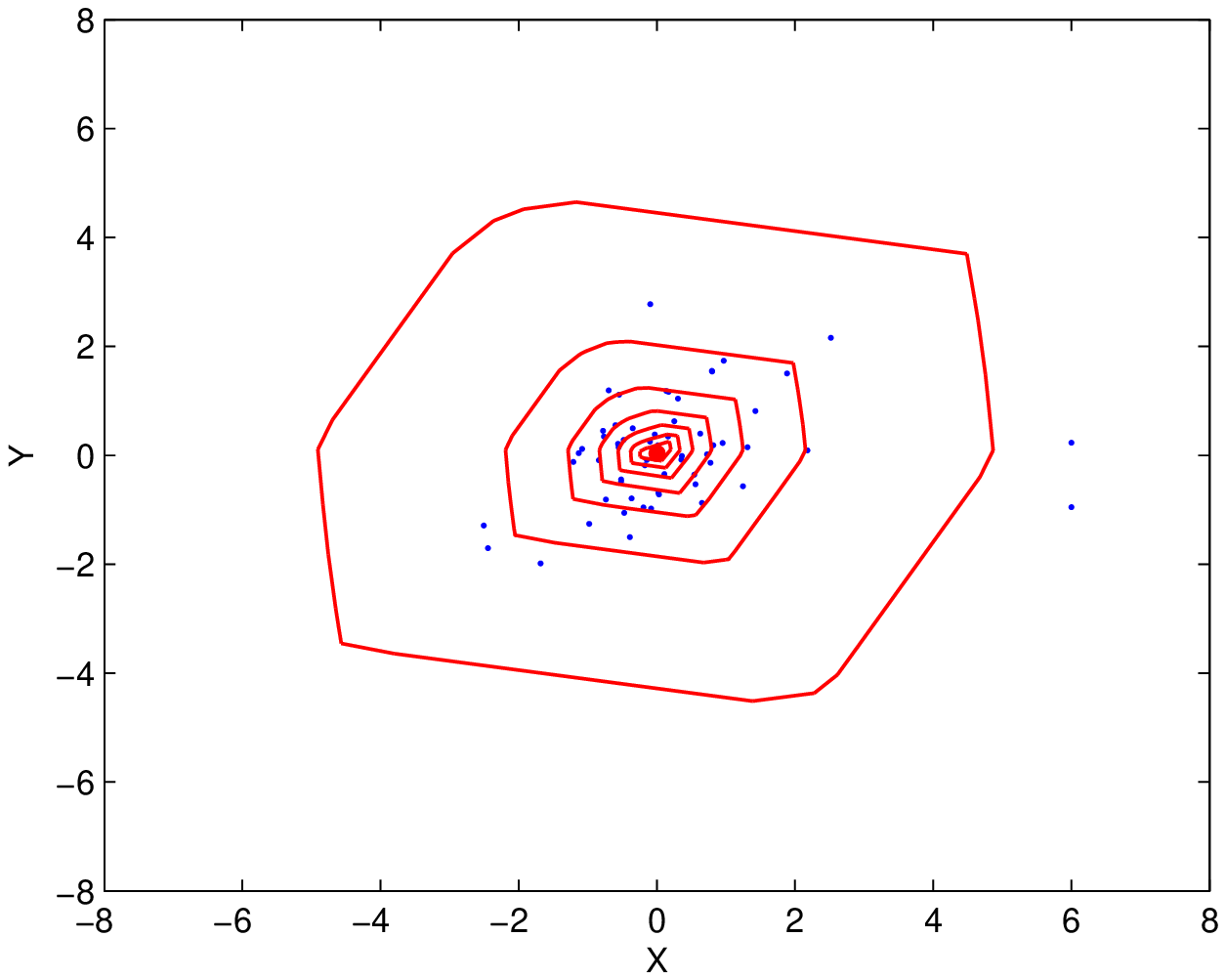}\vspace*{-5mm}
    \caption{Sample version $PC(\alpha, \mathcal{X}^{n})$ for $\alpha = 0.1, 0.2, \cdots, 0.7$ (from out to center). Here the most inner big point denotes the projection median.}
    \label{Norm1SV60}
    \end{center} \vspace*{-8mm}
\end{figure}

\begin{figure}[H] %\vspace*{-8mm}
    \vspace*{-8mm}
    \begin{center}
    \includegraphics[angle=0,width=3.5in]{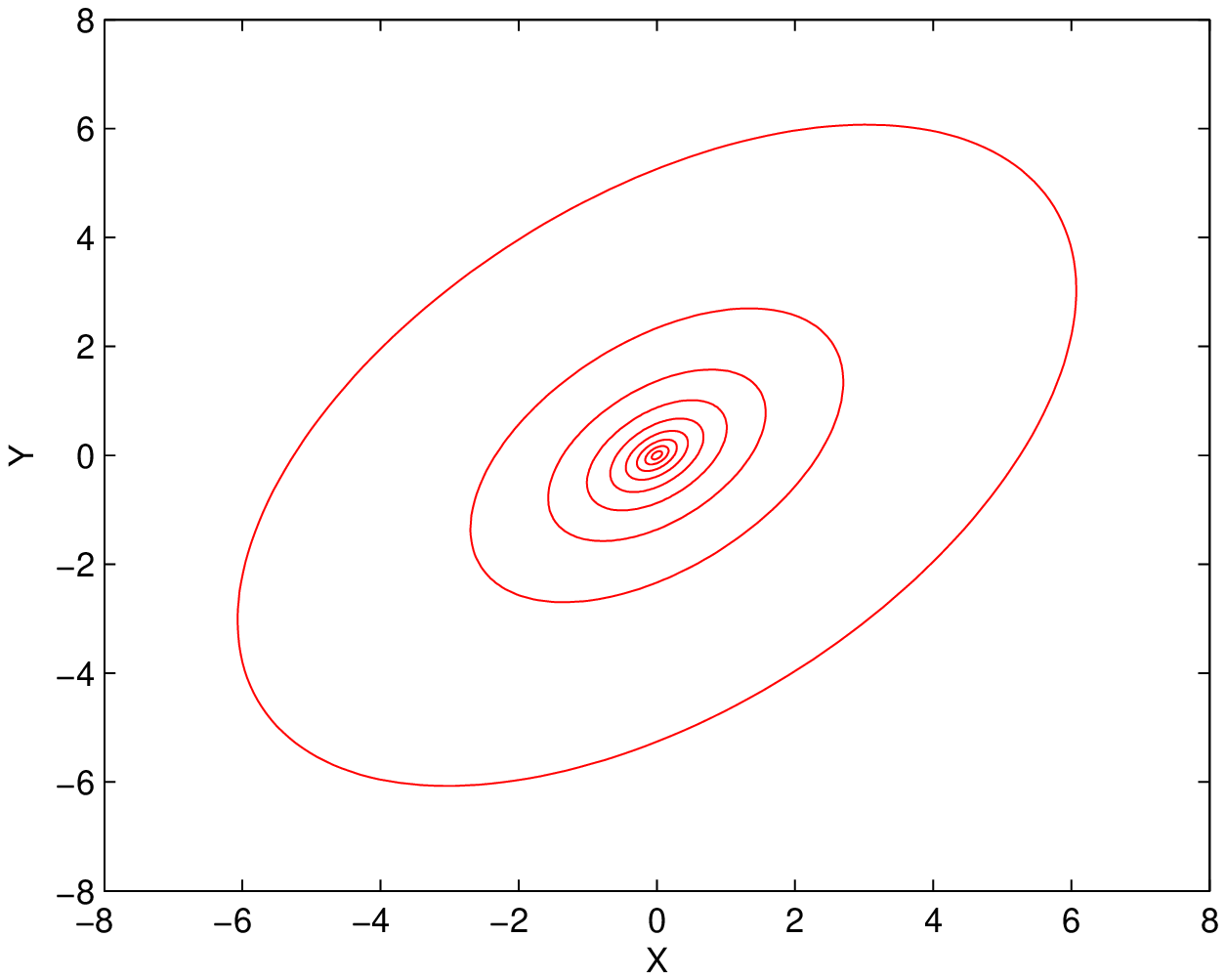}%
    \caption{Population version $PC(\alpha, X)$ for Example 2 with $\alpha = 0.1, 0.2, \cdots, 0.9$ (from out to center)}
    \label{Norm2PV400}
    \includegraphics[angle=0,width=3.5in]{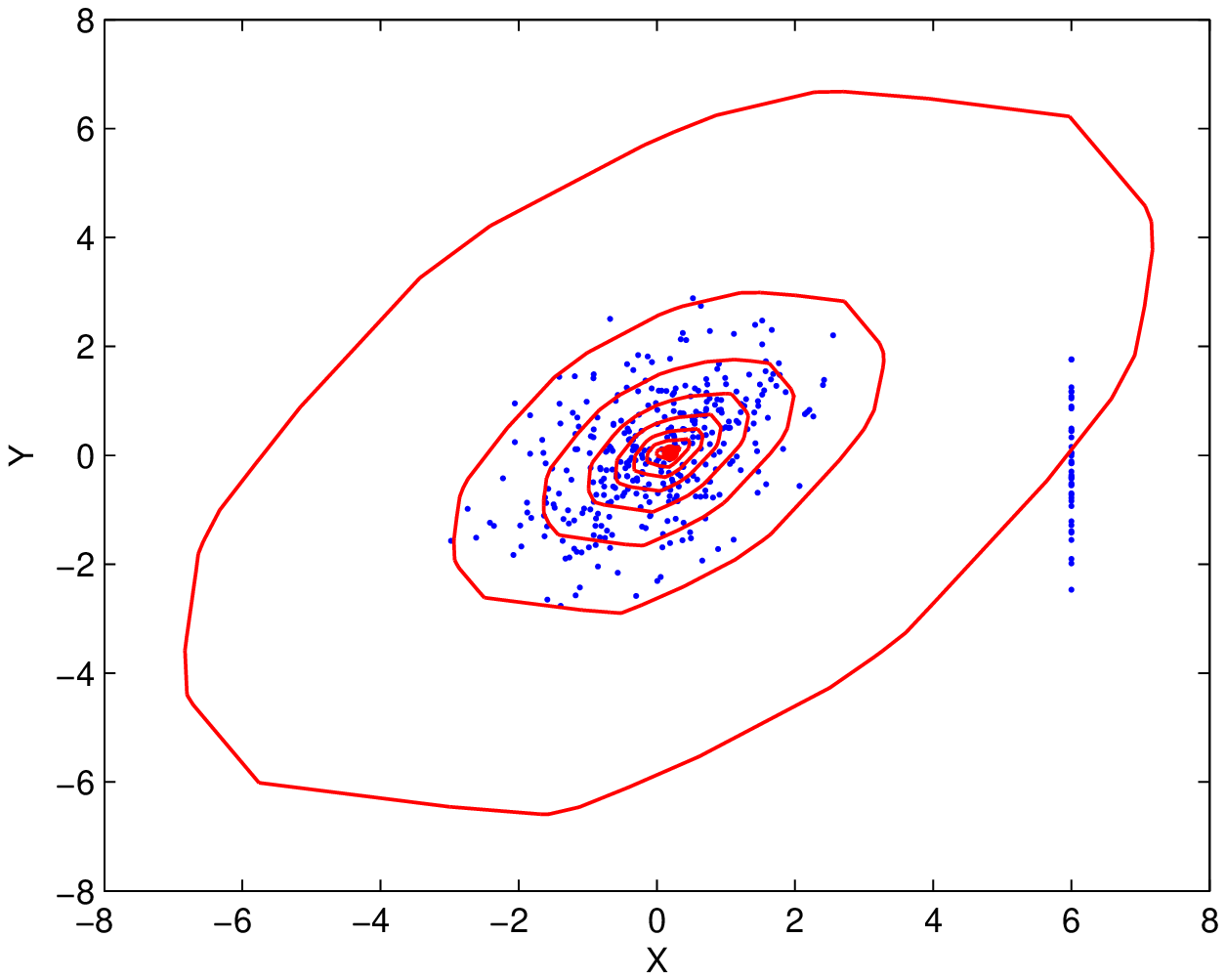}\vspace*{-5mm}
    \caption{Sample version $PC(\alpha, \mathcal{X}^{n})$ for $\alpha = 0.1, 0.2, \cdots, 0.8$ (from out to center). Similar to Figure \ref{Norm1SV60}, the most inner big point is also the computed projection median.}
    \label{Norm2SV400}
    \end{center} %\vspace*{-7mm}
\end{figure}

For the sake of comparison, the population versions of $PC(\alpha, X)$ corresponding to these two examples are also provided here, and plotted according to the formula
$$
    (x_{1}, x_{2}) \times \Sigma^{-1} \times {x_{1} \choose x_{2}} = \frac{C_{N}^{2}(1 - \alpha)^{2}}{\alpha^{2}},
$$
which was developed in \cite{Zuo2003}. Here $C_{N} = \Phi^{-1}(\frac{3}{4}) \approx 0.6744898$, $\Sigma$
denotes the covariance matrix of normally distribution $X$, namely, $I_{2}$ in example 1 and $\Sigma_{0}$ in example 2.
The population versions $PC(\alpha, X)$ of example 1 and 2  are given in Figure \ref{Norm1PV60} and \ref{Norm2PV400},
respectively, while the sample versions $PC(\alpha, \mathcal{X}^{n})$  are plotted in Figure \ref{Norm1SV60} and
\ref{Norm2SV400}, respectively.
\medskip

Comparing the figures of sample versions with those of population, it is ready to see that

\begin{itemize} \vspace*{-1mm}

\item{} The sample versions $PC(\alpha, \mathcal{X}^{n})$ are roughly elliptical,  similar to the shapes of the population versions $PC(\alpha, X)$. Furthermore, it is remarkable that they are resistant to the notorious vertical outliers in Figure \ref{Norm1SV60} and \ref{Norm2SV400}. These results confirm that $PC(\alpha, \mathcal{X}^{n})$ are very robust, and could capture the feature of $PC(\alpha, X)$ \citep{Zuo2003}, even when there are a few outliers.
\vspace*{-1mm}

\item{} It is very interesting to note that the shapes of the projection depth contours are also polylateral due to the properties of the function $\max_{1\leq i\leq M} g_{i}(x)$, similar to that of halfspace depth contours \citep{RutRou1996}. On the other hand, unlike halfspace depth contours, $PC(\alpha, \mathcal{X}^{n})$ does not need to pass through the observations.

\end{itemize}

\newpage

\begin{figure}[H] %\vspace*{-8mm}
    \begin{center}
    \includegraphics[angle=0,width=3.5in]{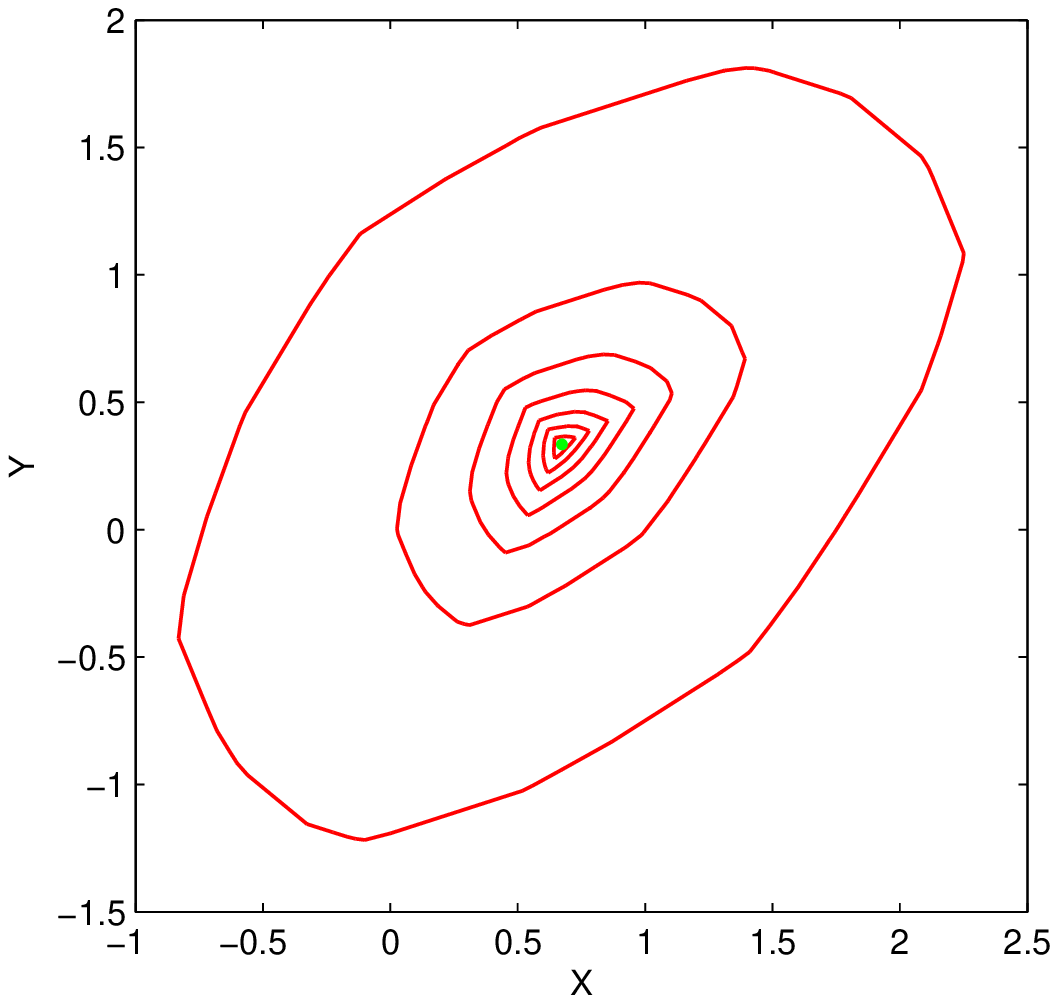}%
    \caption{Projection depth contours of the uniform distribution over the triangle formed by vertexes: $(0, 0)$, $(0, 1)$ and $(1, 1)$, where $\alpha = 0.1, 0.2, \cdots, 0.7$ (from out to center).}
    \label{Unif01SV2500}
    \includegraphics[angle=0,width=3.5in]{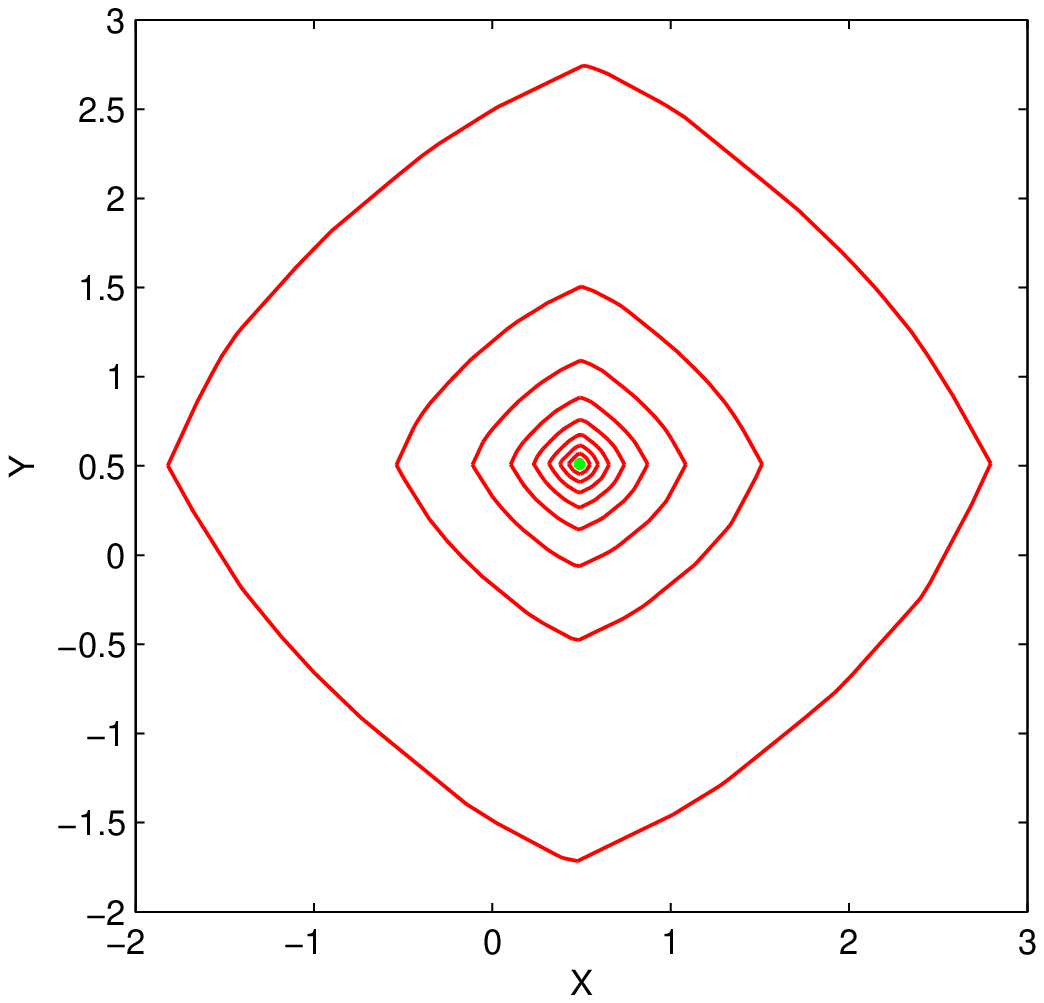}\vspace*{-5mm}
    \caption{Projection depth contours of the uniform distribution over $[0, 1]\times [0, 1]$, where $\alpha = 0.1, 0.2, \cdots, 0.9$ (from out to center).}
    \label{Unif0101SV2500}
    \end{center} %\vspace*{-7mm}
\end{figure}

\begin{figure}[H] %\vspace*{-8mm}
    \begin{center}
    \includegraphics[angle=0,width=3.5in]{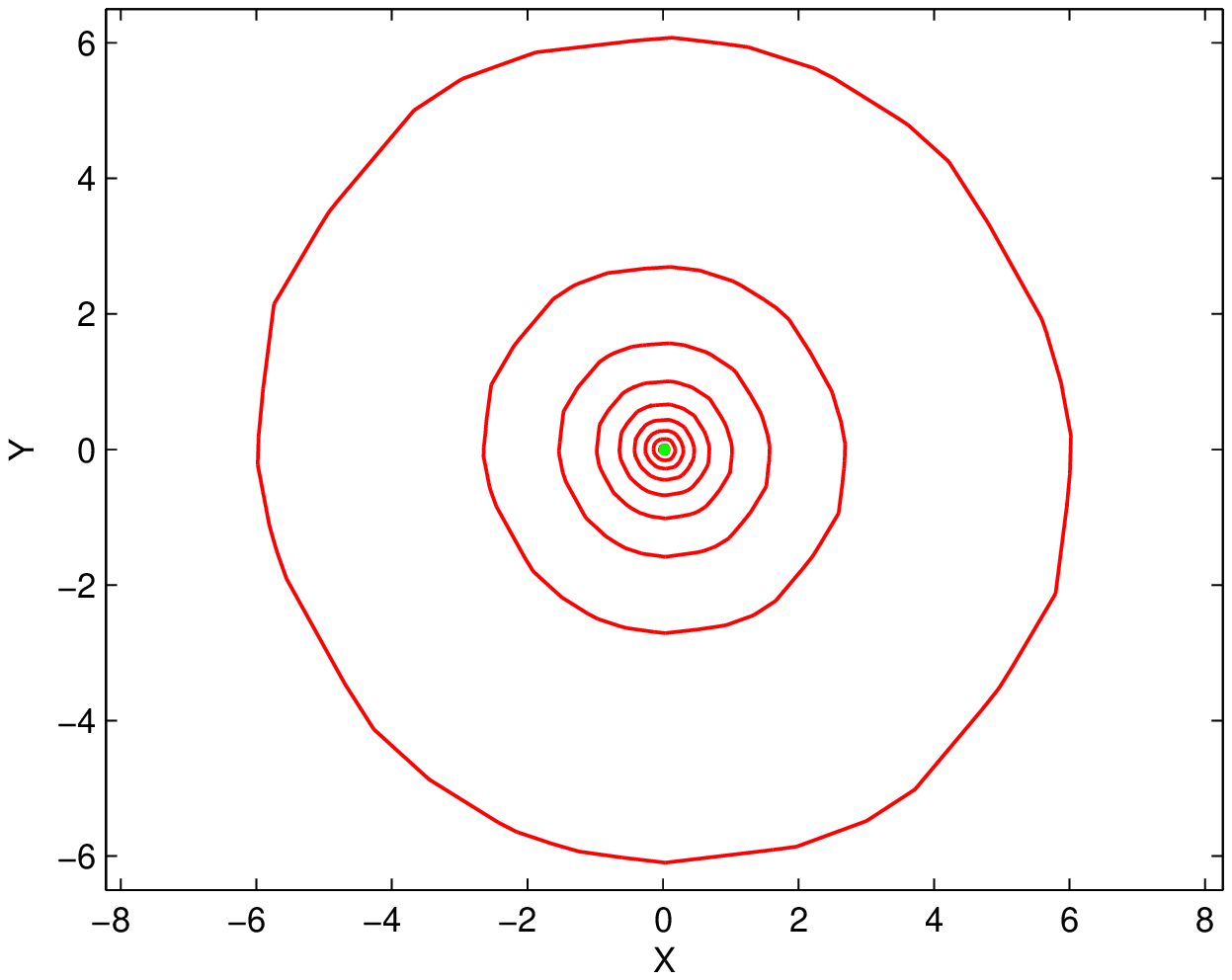}%
    \caption{Projection depth contours of the bivariate standard normal distribution $N(0, I_{2})$, where $\alpha = 0.1, 0.2, \cdots, 0.9$ (from out to center).}
    \label{NormSV2500}
    \includegraphics[angle=0,width=3.5in]{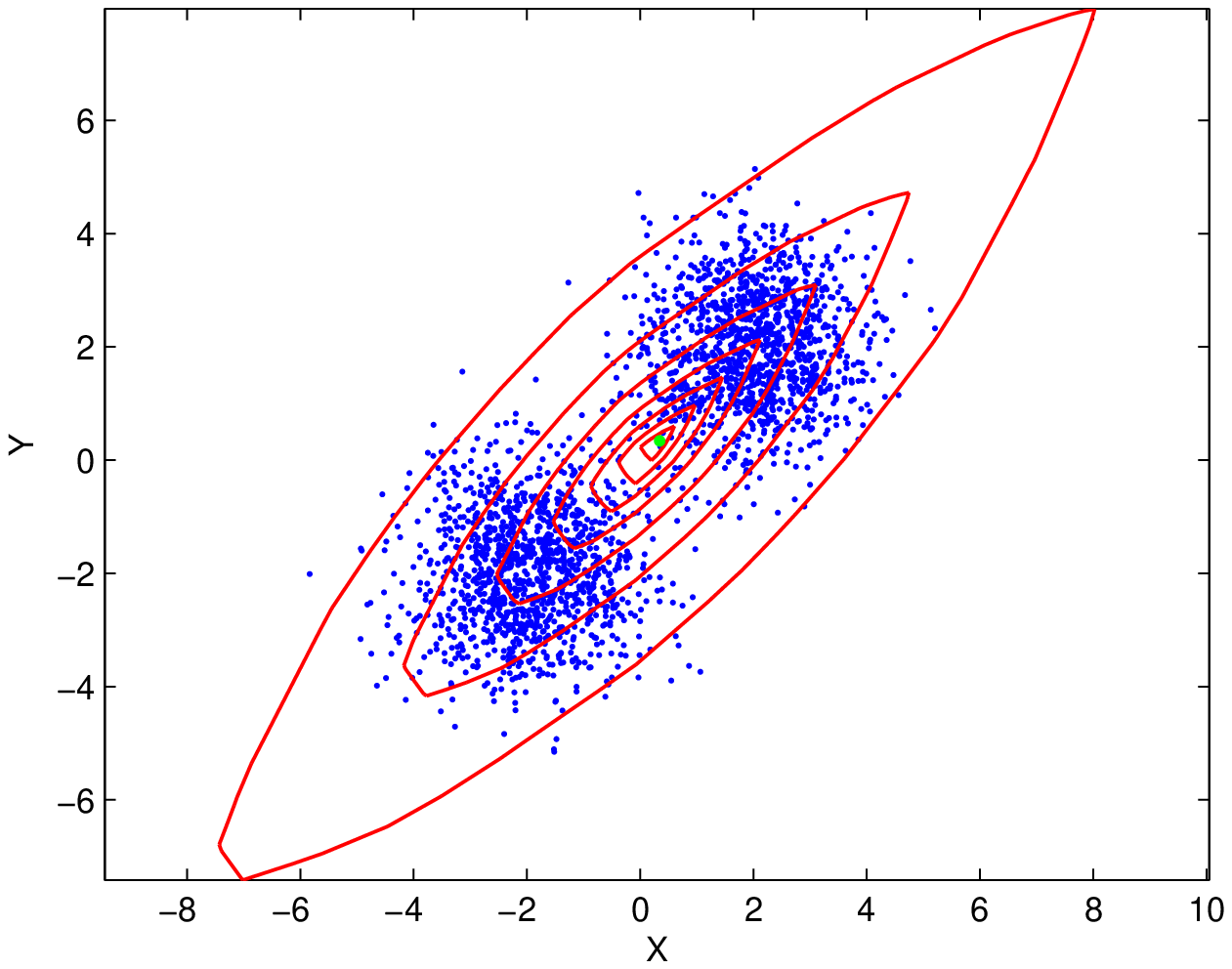}\vspace*{-5mm}
    \caption{Projection depth contours of the random vector $\delta X + (1 - \delta) Y$, where $\alpha = 0.2, 0.3, \cdots, 0.9$ (from out to center).}
    \label{NormPlusNormSV2500}
    \end{center} %\vspace*{-7mm}
\end{figure}

Furthermore, to gain more information about the shape of the projection depth contours, we also provide some other examples in the following. The sample size we used is 2500. Here Figure \ref{Unif01SV2500} reports the projection depth contours $PC(\alpha, \mathcal{X}^{n})$ corresponding to the uniform distribution over the triangle with its vertexes being $(0, 0)$, $(0, 1)$ and $(1, 1)$. Figure \ref{Unif0101SV2500} corresponds to the uniform distribution over region $[0, 1]\times [0, 1]$. Figure \ref{NormSV2500} gives the contours of the normal distribution $N(0, I_{2})$. While Figure \ref{NormPlusNormSV2500} provides the contours of $\delta X + (1 - \delta) Y$, where $X\sim N(\mu_{1}, I_{2})$, $Y \sim N(\mu_{2}, I_{2})$, $\mu_{1} = (-2, -2)$, $\mu_{2} = (2, 2)$, and $\delta$ is a discrete random variable with probability 0.5 taking value 1 and 0.5 taking value 0. Here $X$, $Y$ and $\delta$ are all independently disturbed. From these figures, we can see that the projection contours are also polylateral and convex.

\vskip 0.1 in
\subsection{Real data example}
\paragraph{}
\vskip 0.1 in \label{realexam}

Here a real data example is presented to illustrate the performance of projection depth contours.
\medskip

Table \ref{Tab41} here is taken from Table 7 of \cite{RouLer1987} (p.57).
Total 28 animals' brain weight (in grams) and body weight (in kilograms) are presented in this table. Before the analysis, logarithmic transformation was taken for the sake of convenience. According to the results of \cite{RouLer1987}, there are five cases considered as outlying, i.e. diplodocus, human, triceratops, rhesus monkey and brachiosaurus. Among them, the most severe cases are diplodocus, triceratops and brachiosaurus. In fact, these three cases are referred to as dinosaurs because they possess a small brain as compared with a heavy body (see Table \ref{Tab41}) and their highly negative residuals can lead to a low slope for the least squares fit. For the remaining two cases, although their actual brain weights are higher than those predicted by the linear model, they are not worse than the three previous cases since they do not obey the same trend as that one followed by the majority of the data.

%\vspace*{-1mm}

\begin{table}[t]
{\scriptsize
\begin{center}
    \caption{Body and Brain Weight for 28 Animals \citep{RouLer1987}.}
    \label{Tab41}
    \begin{tabular}{c lp{2.8cm} r r}
    \toprule
        %\multicolumn{4}{c}{$\rho$} \\
                        & Index &           & Body Weight & Brain Weight  \\[0.8ex]
                        & $i$   &  Species  & $X_{i}$    & $Y_{i}$        \\[0.8ex]
        \midrule
                         &1     &Mountain beaver    &1.350       &8.100        \\[0.8ex]
                         &2     &Cow                &465.000     &423.000      \\[0.8ex]
                         &3     &Gray wolf          &36.330      &119.500      \\[0.8ex]
                         &4     &Goat               &27.660      &115.000      \\[0.8ex]
                         &5     &Guinea pig         &1.040       &5.500        \\[0.8ex]
                         &6     &Diplodocus         &11700.000   &50.000       \\[0.8ex]
                         &7     &Asian elephant     &2547.000    &4603.000     \\[0.8ex]
                         &8     &Donkey             &187.100     &419.000      \\[0.8ex]
                         &9     &Horse              &521.000     &655.000      \\[0.8ex]
                         &10    &Potar monkey       &10.000      &115.000      \\[0.8ex]
                         &11    &Cat                &3.300       &25.600       \\[0.8ex]
                         &12    &Giraffe            &529.000     &680.000      \\[0.8ex]
                         &13    &Gorilla            &207.000     &406.000      \\[0.8ex]
                         &14    &Human              &62.000      &1320.000     \\[0.8ex]
                         &15    &African  elephant  &6654.000    &5712.000     \\[0.8ex]
                         &16    &Triceratops        &9400.000    &70.000       \\[0.8ex]
                         &17    &Rhesus monkey      &6.800       &179.000      \\[0.8ex]
                         &18    &Kangaroo           &35.000      &56.000       \\[0.8ex]
                         &19    &Hamster            &0.120       &1.000        \\[0.8ex]
                         &20    &Mouse              &0.023       &0.400        \\[0.8ex]
                         &21    &Rabbit             &2.500       &12.100       \\[0.8ex]
                         &22    &Sheep              &55.500      &175.000      \\[0.8ex]
                         &23    &Jaguar             &100.000     &157.000      \\[0.8ex]
                         &24    &Chimpanzee         &52.160      &440.000      \\[0.8ex]
                         &25    &Brachiosaurus      &87000.000   &154.500      \\[0.8ex]
                         &26    &Rat                &0.280       &1.900        \\[0.8ex]
                         &27    &Mole               &0.122       &3.000        \\[0.8ex]
                         &28    &Pig                &192.000     &180.000      \\[0.8ex]
        \bottomrule
    \end{tabular}
\end{center}}
\end{table}

%\vspace*{-5mm}

\begin{figure}[H]
\vspace*{0mm}
    \begin{center}
    \includegraphics[angle=0,width=3.65in]{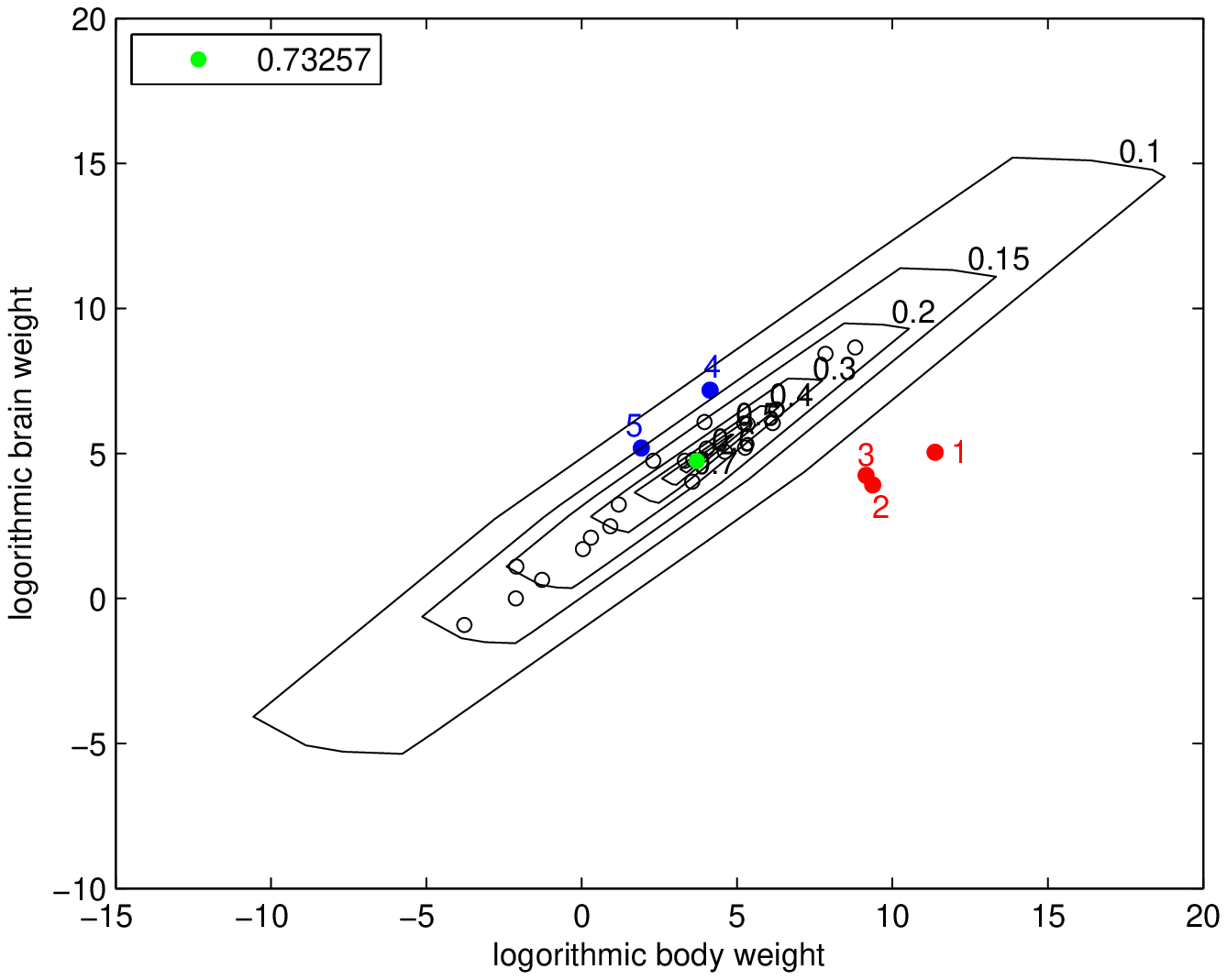}
    \vspace*{-5mm}
    \caption{Projection depth contours with $\alpha = 0.7, 0.6, \cdots, 0.2, 0.15, 0.1$ (from center to out).}
    \label{Figure008}
    \vspace*{-4mm}
    \includegraphics[angle=0,width=3.5in]{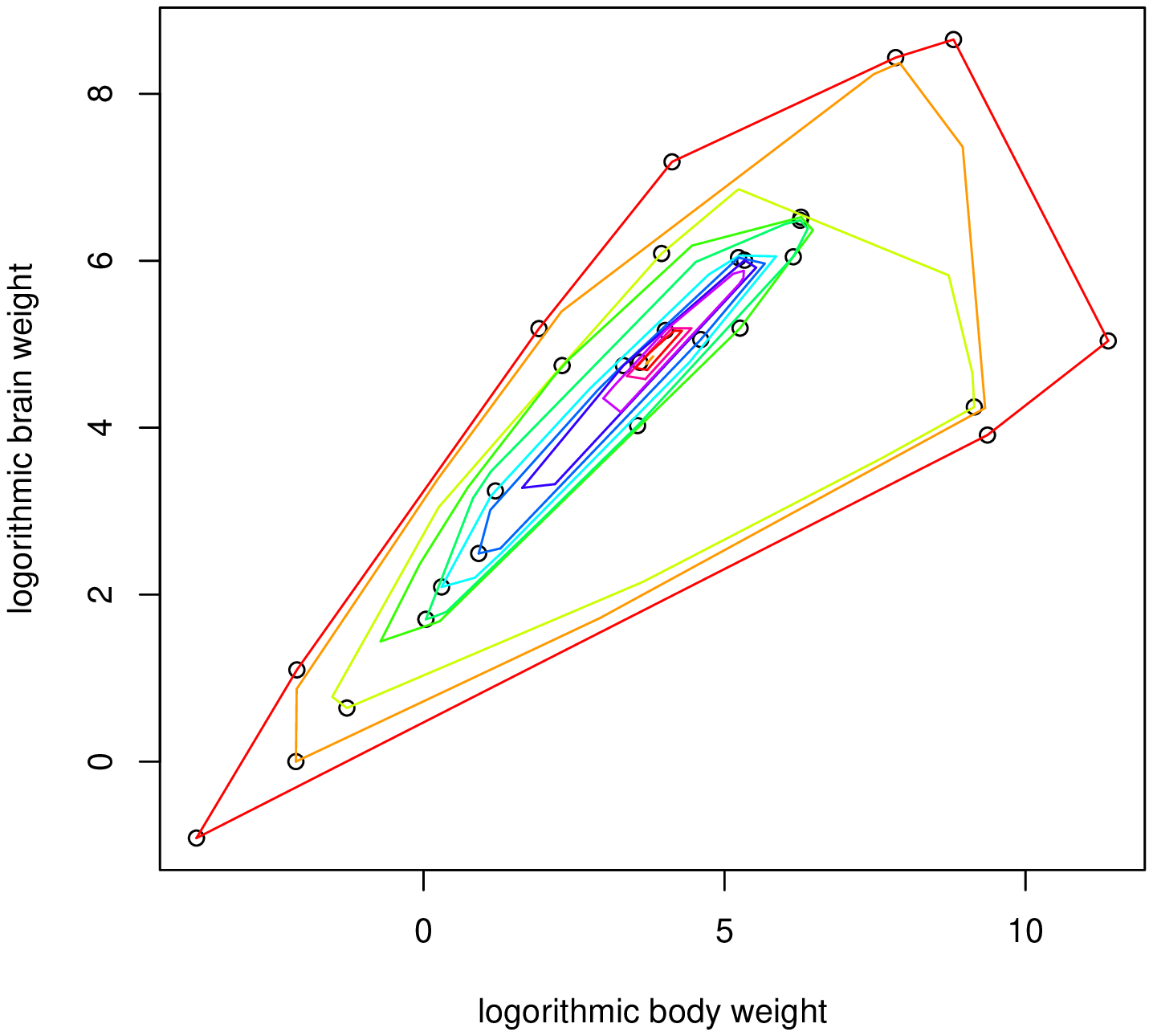}
    \vspace*{-4mm}
    \caption{Halfspace depth contours \citep{RutRou1996}.}
    \label{Figure009}
    \end{center}
\end{figure}

\smallskip
We plot the projection depth contours in Figure \ref{Figure008}, where the green point is the
projection median with depth value 0.73257, five labeled points denote the outliers mentioned above with the points
1-3 corresponding to the case of diplodocus, triceratops and brachiosaurus and 4-5 corresponding to those of human and
rhesus monkey. 8 contours are plotted there. From Figure \ref{Figure008}, we can see that all of these three
dinosaurs lie outside the contour 0.1, while points 4-5 lie between the contours 0.1 and 0.15. These results are consistent with those of \cite{RouLer1987}, implying that projection depth contours can capture the structures of the objective data and identify outliers.
\medskip

Furthermore, it is worth mentioning that %from Figure \ref{Figure008}, we can see that
the shape of these plotted contours is not affected by a few atypical points at the outskirts of the cloud, namely,
both the inner and outer depth contours are roughly elliptical, unlike those of halfspace depth contours (see Figure
\ref{Figure009}) \citep{RutRou1996}.  This is the most outstanding difference between projection and halfspace depth
contours, confirming the more robustness property of projection depth and its contours (see \cite{Zuo2004}).
% See Figure \ref{Figure008} and Figure \ref{Figure009} for comparison.

\newpage

\vskip 0.1 in
\section*{Acknowledgements}
\paragraph{}
\vskip 0.1 in

This work was done during Xiaohui Liu's visit to the Department of Statistics and Probability at Michigan State
University as a joint PhD student. He thanks his co-advisor Professor Yijun Zuo for stimulating discussions and
insightful comments and suggestions and the department for providing excellent studying and working condition. Finally, the authors would like to thank Professor James Stapleton, two anonymous referees, an associate editor and the chief editor Stanley Azen of Computational Statistics and Data Analysis for their careful reading of the first version of this paper. Their constructive comments led to substantial improvements to the manuscript. %This article was partially supported by the National Natural Science of China (No. 61070236).

\vskip 0.1 in
\section*{Appendix  { {\sc Proofs of main results}}}
\paragraph{}
\vskip 0.1 in

%In this appendix, we will give the proofs of the main results.
\medskip

\textbf{Proof of Proposition 1}. Here, without loss of generality, we prove only the odd $n$ case. Note that, for any $u \in \mathcal{Q}_{k}$, we have  
$$
	u^{T}X_{i_{1}} \leq u^{T}X_{i_{2}} \leq \cdots \leq u^{T}X_{i_{m}} \leq \cdots \leq u^{T}X_{i_{n}} 
$$
according to the definition of $\mathcal{Q}_{k}$. This implies that 
$$
    |u^{T}(X_{i} - X_{i_{k, m}})| = \left\{
    \begin{array}{ll}
        -u^{T}(X_{i} - X_{i_{k, m}}),  & \quad \text{if } i \in \{i_{k, 1}, i_{k, 2}, \cdots, i_{k, m -1}\},\\
        u^{T}(X_{i} - X_{i_{k, m}}), & \quad \text{if } i \in \{i_{k, m}, i_{k, m + 1}, \cdots, i_{k, n}\}.
    \end{array}\right.
$$
That is, we can remove the absolute value signs of $Y_{j_{l}}$ based on the order information existing in the permutation $(i_{k, 1}, i_{k, 2}, \cdots, i_{k, n})$. Therefore, for any $u \in \mathcal{Q}_{k}$, \eqref{TmpEq} can be further simplified to 
$$
	Q(u, x, \mathcal{X}^{n}) = \frac{c_{k}^{T}u}{d_{k}^{T}u}
$$
where $c_{k} = x - X_{i_{k}, m}$ and $d_{k} = s_{k}(j_{k, m}) \cdot (X_{j_{k}, m} - X_{i_{k}, m})$ $(1\leq k\leq N)$, with
$$
    s_{k}(i) = \left\{
    \begin{array}{ll}
        -1,  & \quad \text{if } i \in \{i_{k, 1}, i_{k, 2}, \cdots, i_{k, m -1}\},\\
        1, & \quad \text{if } i \in \{i_{k, m}, i_{k, m + 1}, \cdots, i_{k, n}\}.
    \end{array}\right.
$$

Next, note that, for any positive $\lambda$ and $z = \lambda u$, it holds that $c_{k}^{\tau}z / d_{k}^{\tau}z =
c_{k}^{\tau}u / d_{k}^{\tau}u$ and $b_{k}^{\tau}z \ge 0$ if $b_{k}^{\tau}u \ge 0$. Then,  \eqref{TmpEq} and constrain condition
$\mathcal{Q}_k$ %\eqref{TemEqn002}
lead to
\begin{equation}
    O_{k}(x, \mathcal{X}^{n}) = \sup_{z} \frac{c_{k}^{\tau} z}{d_{k}^{\tau} w}
    \label{LinearProOrig}
\end{equation}
subject to
$$
	A_{k} z \ge 0
$$
$1\leq k\leq N$, with $A_{k} = {A_{k1} \choose A_{k2}}$, where
$$
    A_{k1} =
    (X_{i_{k, 2}}^{\tau} - X_{i_{k, 1}}^{\tau},
    X_{i_{k, 3}}^{\tau} - X_{i_{k, 2}}^{\tau},
    \cdots,
    X_{i_{k, n}}^{\tau} - X_{i_{k, n - 1}}^{\tau})^{\tau}
$$
and
$$
    A_{k2} = \left(
    \begin{array}{c}
        s_{k}(j_{k, 2}) \cdot (X_{j_{k, 2}}^{\tau} - X_{i_{k, m}}^{\tau})
            - s_{k}(j_{k, 1}) \cdot (X_{j_{k, 1}}^{\tau} - X_{i_{k, m}}^{\tau})\\
        s_{k}(j_{k, 3}) \cdot (X_{j_{k, 3}}^{\tau} - X_{i_{k, m}}^{\tau})
            - s_{k}(j_{k, 2}) \cdot (X_{j_{k, 2}}^{\tau} - X_{i_{k, m}}^{\tau})\\
        \vdots\\
        s_{k}(j_{k, n}) \cdot (X_{j_{k, n}}^{\tau} - X_{i_{k, m}}^{\tau})
            - s_{k}(j_{k, n - 1}) \cdot (X_{j_{k, n-1}}^{\tau} - X_{i_{k, m}}^{\tau})
    \end{array}
    \right).
$$

This completes the proof of Proposition 1.
\bigskip

%\textbf{{Proof of Theorem 2}}\\

%Let $\mathcal{V}$ be the set of the minimizers of the outlyingness function $O(x, \mathcal{X}^{n})$. If $\mathcal{V}$ has inner point, we then will seek a contradiction. 
%\medskip

%Assume that $x_{0}$ is a inner point of $\mathcal{V}$. By definition, there must exist a small neighborhood $\mathcal{U}$ of $x_{0}$ such that $x_{0} \in \mathcal{U} \subset \mathcal{V}$. Then, it is easy show that, for any $x \in \mathcal{U}$, we have $O(x, \mathcal{X}^{n}) = t_{0} = \min_{y} O(y, \mathcal{X}^{n})$, and therefore
%\begin{eqnarray}
%\label{deriOfOF}
%	\frac{dO(x, \mathcal{X}^{n})}{dx} \equiv 0.
%\end{eqnarray}

%Let $m_{0}$ be the number of functions, say $g_{i_{1}}(x), g_{i_{2}}(x), \cdots, g_{i_{m_{0}}}(x)$, such that  
%$$
%	g_{i_{k}}(x_{0}) = \max_{1 \leq j \leq M} g_{j}(x) = t_{0}, \quad k = 1, \cdots, m_{0}.
%$$
%Now, let's check the value of $m_{0}$. If $m_{0} = 1$, we have $\frac{dO(x, \mathcal{X}^{n})}{dx} = \textbf{a}_{i_{1}} \neq 0$, then \eqref{deriOfOF} is violated. On the other hand, if $m_{0} \ge 2$, the derivative of function $O(x, \mathcal{X}^{n})$ at point $x_{0}$ does not exist, which also leads to a contradiction. 
%\medskip 

%This completes the proof of Theorem 2.

%\bigskip

\end{document}